\documentclass[12pt]{article}
\usepackage{epsfig}
\begin{document}
\baselineskip 21pt

\begin{center}
 {\large\bf Spin effects in $B_c\to X_{c\bar c}\pi(\rho)$ decays.}
\vspace{4mm}

{\sl Saleev V.A.}\footnote
{Email: saleev@ssu.samara.ru}\\[2mm]
Samara State University, 443011, Samara\\
and\\
Samara Municipal Nayanova
University, 443001, Samara\\
Russia
\end{center}

\begin{abstract}
The two-particle hadronic decays of $B_c$ meson into S-wave
and P-wave charmonium states $X_{c\bar c}$ are considered in the
framework of hard gluon exchange model. It is shown that decay width
of $B_c$ meson into S-wave charmonium states doubly
exceeds one for P-wave states.
In compare with the previous estimations we predict
the enhancement by 8 \% the branching ratio for the
$B_c$ decays into $J/\psi$ in the two-particle hadronic decays
via the contribution of the radiative decays of P-wave states
in the cascade processes.
\end{abstract}

\section{Intoduction}
The first observation of $B_c$ meson at FNAL \cite{a1} motivates new
theoretical study in the field of heavy quarkonium physics. The unique
properties of $B_c$ meson, containing heavy quarks of different
flavours, may be studied, first of all, via decay modes with
$J/\psi$ meson in the final state.
We consider here two-particle hadronic decays of $B_c$ meson into S-wave

and P-wave charmonium states
\begin{equation}
      B_c\to X_{c\bar c}+\pi(\rho),
\end{equation}
where $X_{c\bar c}=\eta_c$ or $J/\psi$  for S-wave states, and
$X_{c\bar c}=h_c$, $\chi_{c0}$, $\chi_{c1}$, $\chi_{c2}$ for P-wave
states. These decays are of considerable interest as a clean signal of
$B_c$ meson production in high energy collisions \cite{n1}.

Because of the large momentum transfer to the spectator quark ($\vert
k^2 \vert \sim m_c^2\gg \Lambda_{QCD}$) in the decays (1.1),
the hard scattering formalism
is more appropriate than the spectator model, which based on the
transition form-factor calculation under the overlapping of the
quarkonium wave functions.
In contrast to the spectator model, the hard scattering formalism
results in the approximate double enhancement of the decay amplitude
for the decays $B_c\to J/\psi(\eta_c)\pi$, as it was found in
\cite{n2,n3}.

\section{The model}

In the model under consideration it is assumed that heavy quarkonium
($B_c$ or $X_{c\bar c}$) is a non-relativistic quark-antiquark system
with small binding energy. Such a way, the $B_c$ meson decay amplitude
is factorized in a hard part which describes the process $\bar b+c\to
\bar c+c+\pi(\rho)$ and an amplitude describing the binding of the
initial
and the final heavy quark pairs into $B_c$ meson and $X_{c\bar c}$
state.

The general covariant formalism for calculating the production and
decay rates of S-wave and P-wave heavy quarkonium in the
non-relativistic expansion was developed some times ago \cite{n4}.
In the leading non-relativistic approximation, the mass of $B_c$ meson
$m_1$ is simply the sum of b-quark  and c-quark masses $m_b+m_c$, and
the mass of $X_{c\bar c}$ is equal $2m_c$. The amplitude for decay the
bound-state ($\bar b c$)
in a state with momentum $p_1$, total angular
momentum $J_1$, total orbital momentum $L_1$ and total spin $S_1$ into
bound-state ($\bar c c$)
in a state with momentum $p_2$, total angular
momentum $J_2$, total orbital momentum $L_2$ and total spin $S_2$
is given by
\begin{eqnarray}
&&A(p_1,p_2)=
\int \frac{d\vec q_1}{(2\pi)^3}\sum_{L_{1z}S_{1z}}
\Psi_{L_{1z}S_{1z}}(\vec q_1)
<L_1L_{1z};S_1S_{1z}|J_1J_{1z}>\times \\
&&\times \int \frac{d\vec q_2}{(2\pi)^3}\sum_{L_{2z}S_{2z}}
\Psi_{L_{2z}S_{2z}}(\vec q_2)
<L_2L_{2z};S_2S_{2z}|J_2J_{2z}>
M(p_1,p_2,q_1,q_2),\nonumber
\end{eqnarray}
where $M(p_1,p_2,q_1,q_2)$ is the hard amplitude which is described
by the diagrams in Fig. 1,2.

%\vspace{1cm}
%\FIGURE{\epsfig{file=fig1a1.eps,width=7cm}%
%}
%\vspace{1cm}
%\FIGURE{\epsfig{file=fig1a1.eps,width=7cm}%
%\caption{Diagrams for the decay $B_c\to X_{c\bar c}\pi(\rho)$.}}
%

To introduce the operators $\Gamma_{SS_z}(p,q)$ which projects the
quark-antiquark pairs onto the bound states with fixed quantum number
up to second order in $q_1$ and $q_2$:
\vspace{5mm}

\begin{eqnarray}
&&\Gamma_{S_1S_{1z}}(p_1,q_1)=\frac{\sqrt{m_1}}{4m_cm_b}
(\frac{m_c}{m_1}\hat p_1-\hat q_1+m_c)\hat A_1
(\frac{m_b}{m_1}\hat p_1+\hat q_1-m_b),
\end{eqnarray}
where $\hat A_1=\gamma_5$ for $S_1=0$ and
$\hat A_1=\hat \varepsilon(S_{1z})$  for $S_1=1$;

\begin{eqnarray}
&&\Gamma^\dagger_{S_2S_{2z}}(p_2,q_2)=\frac{\sqrt{m_2}}{4m_c^2}
(\frac{m_c}{m_2}\hat p_2+\hat q_2-m_c)\hat A_2
(\frac{m_c}{m_2}\hat p_2-\hat q_2+m_c),
\end{eqnarray}
where $\hat A_2=\gamma_5$ for $S_2=0$,
$\hat A_2=\hat \varepsilon(S_{2z})$  for $S_2=1$ and
$\varepsilon(S_{1z,2z})$ are  spin-one polarization four-vectors.

Using projection operators (2.2) and (2.3) the hard amplitude
$M(p_1,p_2,q_1,q_2)$ may be presented as follows:
\begin{equation}
M(p_1,p_2,q_1,q_2)=\mbox{Tr}\left [
\Gamma^\dagger
(p_2,q_2)\gamma^{\beta}\Gamma(p_1,q_1)\cal{O}_{\beta}\right ],
\end{equation}
where for decay with $\pi$ meson in final state
\begin{eqnarray}
{\cal O}_{\beta}&&={\cal O}^1_{\beta}+{\cal O}^2_{\beta},\\
{\cal O}^1_{\beta}&&=\frac{G_F}{\sqrt{2}}\frac{16\pi\alpha_s}{3}V_{bc}f_{\pi}a_1\hat p_3 (1-\gamma_5)
\left (\frac{-\hat x_1+m_c}{x_1^2-m_c^2}
\right )\frac{\gamma_{\beta}}{k^2},\\
{\cal
O}^2_{\beta}&&=\frac{G_F}{\sqrt{2}}\frac{16\pi\alpha_s}{3}V_{bc}f_{\pi}a_1
\frac{\gamma_{\beta}}{k^2}\left (\frac{-\hat x_2+m_b}{x_2^2-m_b^2}
\right )\hat p_3(1-\gamma_5),
\end{eqnarray}
and
$$ \hat x_1=\frac{m_b}{m_1}\hat p_1+\hat q_1-\hat p_3, \quad
\hat x_2=\frac{m_c}{m_2}\hat p_2+\hat q_2+\hat p_3$$
$$ \hat k=\frac{m_c}{m_2}\hat p_2-\frac{m_c}{m_1}\hat p_1-\hat q_1+
\hat q_2.$$
The factor $a_1$ comes from hard gluon corrections to the four-fermion
effective vertex.
Since $q_1/m_1$ and $q_2/m_2$ are small quantities, we can expand
$M(p_1,p_2,q_1,q_2)$ around $q_1=q_2=0$ in a Taylor expansion:
\begin{eqnarray}
&&M(p_1,p_2,q_1,q_2)=M(p_1,p_2,0,0)+
q_{1\alpha}\frac{\partial M}{\partial q_{1\alpha}}|_{q_{1,2}=0}+
q_{2\alpha}\frac{\partial M}{\partial
q_{2\alpha}}|_{q_{1,2}=0}+\nonumber\\
&&+\frac{1}{2}q_{1\alpha}q_{1\beta}
\frac{\partial^2 M}{\partial q_{1\alpha}
\partial q_{2\beta}}|_{q_{1,2}=0}+...
\end{eqnarray}
Here the each term correspond to quantum numbers
$L_1=L_2=0$, $L_1=1$ and $L_2=0$, $L_1=0$ and $L_2=1$, and so forth.
Thus for the S-wave ($L=0$) and P-wave ($L=1$) states, the amplitude
$A(p_1,p_2)$ will depend on the quarkonium radial wave-functions
through the following relations:
\begin{eqnarray}
&&\int \frac{d\vec q}{(2\pi)^3}\Psi_{00}(\vec q)
=\frac{R_s(0)}{\sqrt{4\pi}},\\
&&\int \frac{d\vec q}{(2\pi)^3}\Psi_{1L_z}(\vec q)q_{\alpha}=
-i\sqrt{\frac{3}{4\pi}}R'_p(0)\varepsilon_{\alpha}(p,L_z),
\end{eqnarray}
where $\varepsilon_{\alpha}(p,L_z)$ is the polarization vector for the
spin-one particle.

In the case of production charmonium state $^1P_1$ one has
\begin{equation}
\sum_{L_{2z}}\varepsilon^{\alpha}(p_2,L_{2z})<1L_{2z},00\vert
1,J_{2z}>=\varepsilon^{\alpha}(p_2,J_{2z}).
\end{equation}
The summation over polarization may be done using following expression
\begin{equation}
\sum_{J_{2z}=-1}^1\varepsilon^{\alpha}(p2,J_{2z})
\varepsilon^{\beta}(p2,J_{2z})={\cal P}^{\alpha\beta}(p_2),
\end{equation}
where
$${\cal
P}^{\alpha\beta}(p_2)=-g^{\alpha\beta}+\frac{p_2^{\alpha}p_2^{\beta}}
{m_2^2}.$$
In the case of production $^3P_J(J=0,1,2)$ states one has
\begin{eqnarray}
&&\sum_{S_{2z},L_{2z}}\varepsilon^{\alpha}(p_2,L_{2z})<1L_{2z},1S_{2z}
\vert J_2,J_{2z}>\varepsilon^{\beta}(S_{2z})=\nonumber\\
&&=\left \{ \begin{array}{l}
\frac{1}{\sqrt{3}}(g^{\alpha\beta}-\frac{p_2^{\alpha}p_2^{\beta}}
{m_2^2})\mbox{ for }J_2=0,\\
\frac{i}{\sqrt{2}m_2}\varepsilon^{\alpha\beta\mu\nu}p_{2\mu}
\varepsilon_{\nu}(p_2,J_{2z})\mbox{ for }J_{2}=1,\\
\varepsilon^{\beta\alpha}(p_2,J_{2z})\mbox{ for } J_2=2.
\end{array}
\right.
\end{eqnarray}
The polarization sums for $J_2=2$ is given by the following
expression \cite{n5}:
\begin{eqnarray}
\sum_{J_{2z}=-2}^2\varepsilon_{\alpha\beta}(p_2,J_{2z})
\varepsilon^*_{\mu\nu}(p_2,J_{2z})&=&
\frac{1}{2}({\cal P}_{\alpha\mu}(p_2){\cal P}_{\beta\nu}(p_2)+
{\cal P}_{\alpha\nu}(p_2){\cal P}_{\beta\mu}(p_2))-\nonumber\\
&&-\frac{1}{3}{\cal P}_{\alpha\beta}(p_2){\cal P}_{\mu\nu}(p_2))
\end{eqnarray}

\section{The results}
Omitting the details of calculations we presented below the results
for decay widths of ground-state $(\bar b c)$ system into different
states of charmonium plus $\pi$ or $\rho$ meson. In the limit of
vanishing
$\pi$ meson mass we obtained the simple analytical formulae:
\begin{eqnarray}
&&\Gamma (B_c\to \psi\pi)=\frac{128}{9\pi}F\frac{|R_2(0)|^2}{m_2^3}
\frac{(1+x)^3}{(1-x)^5},\\
&&\Gamma (B_c\to \eta_c\pi)=\frac{32}{9\pi}F\frac{|R_2(0)|^2}{m_2^3}
\frac{(1+x)^3}{(1-x)^5}(x^2-2x+3)^2,\\
&&\Gamma (B_c\to h_c\pi)=\frac{128}{3\pi}F\frac{|R'_2(0)|^2}{m_2^5}
\frac{(1+x)^3}{(1-x)^7}(x^2-x+2)^2,\\
&&\Gamma (B_c\to
\chi_{c0}\pi)=\frac{128}{9\pi}F\frac{|R'_2(0)|^2}{m_2^5}
\frac{(1+x)^3}{(1-x)^7}(3x^3-12x^2+14x-7)^2,\\
&&\Gamma (B_c\to
\chi_{c1}\pi)=\frac{256}{3\pi}F\frac{|R'_2(0)|^2}{m_2^5}
\frac{(1+x)^3}{(1-x)^5}(x^2-x-1)^2,\\
&&\Gamma (B_c\to
\chi_{c2}\pi)=\frac{256}{9\pi}F\frac{|R'_2(0)|^2}{m_2^5}
\frac{(1+x)^5}{(1-x)^7},
\end{eqnarray}
where
$$x=\frac{m_2}{m_1}, \mbox{ and }
F=\alpha_s^2G_F^2V_{bc}^2f_{\pi}^2|R_1(0)|^2a_1^2.$$
To perform the numerical calculations we use following set
of parameters: $G_F=1.166\times 10^{-5}$ GeV$^{-2}$,
$\alpha_s=0.33$, $V_{bc}=0.04$,
$f_{\pi}=0.13$  GeV,
$m_{\pi}=0.14$  GeV,
$m_{B_c}=6.3 $  Gev,
$m_{\psi}=3.1$ Gev, $m_{\eta_c}=2.98$  Gev,
$m_{h_c}=3.5$ Gev,
$m_{\chi_{c0}}=3.4$ Gev, $m_{\chi_{c1}}=3.5$ Gev,
$m_{\chi_{c2}}=3.55$  Gev,
$|R_{s1}(0)|^2=1.27$  GeV$^3$,
$|R_{s2}(0)|^2=0.94$ GeV$^3$,
$|R'_{p2}(0)|^2=0.08$ GeV$^5$.

With above mentioned set of parameters one gets the following result
\begin{equation}
    \Gamma (B_c\to J/\psi + \pi)=7.5 \times 10^{-15}a_1^2\mbox{ GeV}.
\end{equation}
The decay widths into different charmonium states plus $\pi$ meson
may be presented through decay width
for $B_c\to J/\psi\pi$ as it is shown in Table 1.

\vspace{4mm}
\begin{center}
\begin{tabular}{|c|c|c|c|}\hline
$X_{c\bar c}$ &
$^{2S+1}X_J$ &
$\Gamma (B_c\to X_{c\bar c}\pi)\over\Gamma (B_c\to J/\psi\pi) $&
$\Gamma (B_c\to X_{c\bar c}\rho)\over\Gamma (B_c\to X_{c\bar c}\pi) $\\
\hline
$J/\psi$ & $^3S_1$ & 1.00 & 4.0 \\ \hline
$\eta_c$ & $^1S_0$ & 1.17 & 3.2 \\ \hline
$h_c$    & $^1P_1$ & 0.50 & 3.7 \\ \hline
$\chi_{c0}$ & $^3P_0$& 0.29 & 3.6 \\ \hline
$\chi_{c1}$ & $^3P_1$& 0.10 & 5.6 \\ \hline
$\chi_{c2}$ & $^3P_2$& 0.28 & 4.3 \\ \hline
\end{tabular}
\end{center}
\vspace{4mm}

\centerline{Table 1.}
\vspace{4mm}

The another source of $J/\psi$ mesons is two-particle decay of $B_c$
meson with $\rho$ meson in the final state: $B_c\to X_{c\bar c}\rho$.
The calculating for the decay widths $\Gamma (B_c\to X_{c\bar c}\rho)$
may be done the same way as for widths $\Gamma (B_c\to X_{c\bar c}\pi)$
using substitution $f_{\pi}\hat p_3\to
m_{\rho}f_{\rho}\hat\varepsilon_3$
in (2.6) and (2.7), where $\varepsilon_3^{\mu}$ is $\rho$ meson
polarization four-vector. Taking into account that $f_{\rho}=0.22$ GeV
and $m_{\rho}=0.77$ GeV, we have obtained the decay widths
$\Gamma (B_c\to X_{c\bar c}\rho)$ which are presented in the Table 1
too as the ratio $\Gamma (B_c\to X_{c\bar c}\rho)/
\Gamma (B_c\to X_{c\bar c}\pi)$.

We found surprisingly large value for the decay width of
$B_c$ meson into P-wave charmonia, which is 50 \% of the decay width
into S-wave states.
Because of $J/\psi$ meson production in the decays of $B_c$ meson is
very suitable process from viewpoint of an experimental study \cite{a1},

it is interesting
to compare the direct $J/\psi$ production
($B_c\to J/\psi\pi(\rho)$) and the cascade $J/\psi$
production rates. The second one comes from radiative decays of the
P-wave states $\chi_{c0},\chi_{c1}$ and $\chi_{c2}$, which have
following branching ratios into $J/\psi$ plus $\gamma$:
$\mbox{Br}(\chi_{c0}\to J/\psi+\gamma)=0.007,$
$\mbox{Br}(\chi_{c1}\to J/\psi+\gamma)=0.27$ and
$\mbox{Br}(\chi_{c2}\to J/\psi+\gamma)=0.14$ \cite{n6}.
Thus we have obtained
\begin{equation}
{\Gamma (B_c\to\chi_{c0,c1,c2}\pi\to J/\psi\gamma)\over
{\Gamma (B_c\to J/\psi\pi)}}=0.068
\end{equation}
and
\begin{equation}
{\Gamma (B_c\to\chi_{c0,c1,c2}\rho\to J/\psi\gamma)\over
{\Gamma (B_c\to J/\psi\rho)}}=0.082
\end{equation}
The ratio for the sum of the $B_c$ meson decay widths into
$J/\psi$ meson plus $\pi$ or $\rho$ meson is equal to
\begin{equation}
{\Gamma (B_c\to J/\psi\pi(\rho),\mbox{cascade})\over
{\Gamma (B_c\to J/\psi\pi(\rho),\mbox{direct}})}=0.08
\end{equation}

%%%%%%%%%%%%%%%%%%%%%%%%%%%%%%%%%%%%%%%%%%%%%%%%%%%%%%%%%%%%%%%%%%%
In compare with the previous estimations \cite{n2,n3} we should expect
the enhancement by 8 \% the branching ratio for the
$B_c$ decays into $J/\psi$ in the two-particle hadronic decays
via the cascade processes. This fact makes the probability of
$B_c$ meson observation through the decays $B_c\to X_{c\bar c}\pi(\rho)$

in the current experiments more real.

The author thanks V.V.~Kiselev and A.K. Likhoded for the valuable
discussion. This work is supported by the Program "Universities of
Russia", Grant 02.01.03 and the Russian Ministry of Education, Grant
98-0-6.2-53.

\end{document}